\documentclass[%
 reprint,
superscriptaddress,
showpacs,
 amsmath,amssymb,
 aps,
 prl,
]{revtex4-1}

\usepackage{graphics,amssymb,amsmath,epsfig,color}
\usepackage{graphicx}

\usepackage{hyperref}
\usepackage{graphicx}
\usepackage{dcolumn}
\usepackage{bm}
\usepackage[usenames, dvipsnames]{xcolor}

\newcommand{\Alfvenic}{Alfv$\acute{\text{e}}$nic}
\newcommand{\Rm}{\mathrm{Rm}}
\renewcommand{\Re}{\mathrm{Re}} 
\newcommand{\Pm}{\mathrm{Pm}}

\usepackage{amssymb} 

\begin{document}

\title{Mechanism for Sequestering Magnetic Energy at Large Scales in Shear-Flow Turbulence}

\author{B.~Tripathi}
\email{btripathi@wisc.edu}
\affiliation{University of Wisconsin-Madison, Madison, Wisconsin 53706, U.S.A.}
\author{A.E.~Fraser}
\email{adfraser@ucsc.edu}
\affiliation{University of California, Santa Cruz, Santa Cruz, California 95064, U.S.A.}
\author{P.W.~Terry}
\email{pwterry@wisc.edu}
\author{E.G.~Zweibel}
\email{zweibel@astro.wisc.edu}
\affiliation{University of Wisconsin-Madison, Madison, Wisconsin 53706, U.S.A.}

\author{M.J.~Pueschel}
\affiliation{Dutch Institute for Fundamental Energy Research, 5612 AJ Eindhoven, The Netherlands}
\affiliation{Eindhoven University of Technology, 5600 MB Eindhoven, The Netherlands}

\date{\today}

\begin{abstract}
Straining of magnetic fields by large-scale shear flow, generally assumed to lead to intensification and generation of small scales, is re-examined in light of the persistent observation of large-scale magnetic fields in astrophysics. It is shown that, in magnetohydrodynamic turbulence, unstable shear flows have the unexpected effect of sequestering magnetic energy at large scales, due to counteracting straining motion of nonlinearly excited large-scale stable eigenmodes. This effect is quantified via dissipation rates, energy transfer rates, and visualizations of magnetic field evolution by artificially removing the stable modes. These analyses show that predictions based upon physics of the linear instability alone miss substantial dynamics, including those of magnetic fluctuations.
\end{abstract}

\maketitle

\begin{figure*}
    \centering
    \includegraphics[width=1\textwidth]{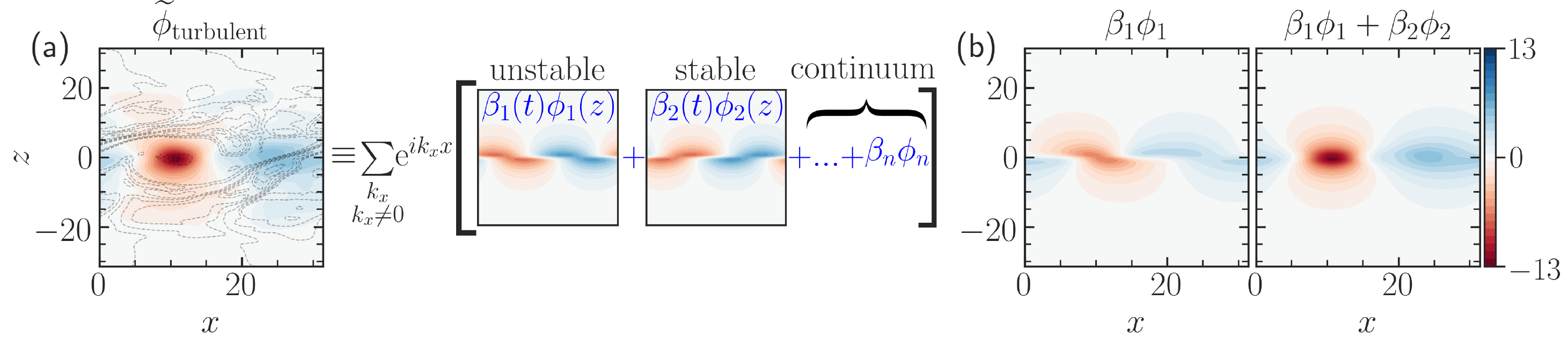}
    \caption{(a) Eigenmode decomposition of a characteristic snapshot of the turbulent flow, represented by the solid- and colored-contour lines of the stream function $\widetilde{\phi}_{\mathrm{turbulent}}$ at time $t=625$, with $\mbox{M}_{\rm A}=10$. The decomposition is based on a complete set: unstable, stable, and continuum eigenmodes. The eigenmodes plotted are for the flow fluctuations at the first Fourier wavenumber. The black dashed-contour lines overplotted on the first image show the turbulent magnetic flux function $\widetilde{\psi}_\mathrm{turbulent}$, whose distortion near the center of the eddy, $(x,z) \approx (10,0)$, is significantly impeded. (b) Reconstruction of the turbulent flow: (left) sum of unstable eigenmodes at each Kelvin-Helmholtz-unstable wavenumber; (right) improvement by adding their conjugate stable eigenmodes. Compare these with full $\widetilde{\phi}_\mathrm{turbulent}$ in (a). The difference between the plots (a) and (b) are shown in the supplementary material. All panels here share the same colorbar.}
    \label{fig:f1}
\end{figure*}

Turbulence is a fundamentally multiscale process in which nonlinear dynamical motions carry energy across scales \cite{johnson2020}. One of the most robust and extensively studied mechanisms for cross-scale energy transfer is straining by shear flow \cite{batchelor1950, batchelor1954}, long recognized for its role in cascades to small scale \cite{townsend1976}. This process is especially active in magnetohydrodynamic (MHD) turbulence \cite{maron2004}. The stretching, squeezing, and folding of magnetic field lines by shear flow is readily discernible in visualizations, and leads to, for example, the enhancement of small-scale resistive dissipation and the intensification of magnetic energy at small scales beyond the viscous cutoff in turbulence with magnetic Prandtl number greater than unity \cite{schekochihin2002, maron2004}. The ubiquity of turbulence and turbulent straining is at odds with the well-established observations of magnetic fields at large scales in the universe (stellar, galactic, and beyond) \cite{neronov2010, kulsrud1999} and part of a conundrum as to how magnetic fields exist at such large scales \cite{brandenburg2005, tobias2013, squire2015}.

In this Letter we report the first observations that magnetic energy is, to a considerable degree, counterintuitively sequestered at large scales in straining by a paradigmatic turbulent shear flow. The result applies to a Kelvin-Helmholtz- (KH-)unstable flow maintained by an external force with a uniform magnetic field that is initially flow-aligned and therefore optimally configured to promote small-scale generation. 

The process that counteracts the small scale generation of magnetic energy, which is quantified in this Letter, can be traced to the nonlinear excitation of large-scale stable linear eigenmodes \cite{hatch2011}. The dispersion relation that is associated with the shear-flow instability present in this study has a stable (damped) root, which acts to return energy and momentum to the mean profile, and which is excited to a high level by the nonlinearity \cite{fraser2017, fraser2021}. The nonlinear excitation of stable modes and its effect on turbulence levels and transport have been studied previously, particularly in the context of fusion-relevant gyroradius-scale turbulence \cite{terry2018, whelan2018, fraser2018, pueschel2016, makwana2014, hatch2013, makwana2012, hatch2011, makwana2011, terry2006}. Recent studies have shown that stable modes are excited in macroscopic shear-flow driven turbulence also and transiently affect momentum transport \cite{fraser2017, fraser2021}. However, critical features of counteracting straining motion of the stable modes, essential for large-scale sequestering of magnetic energy, have only come to light in the study described here.

To whit, stable-mode effects have been linked to small-scale dissipation \cite{hatch2013}, suggesting that they affect turbulence only if there is an increase in entropy. However, we show here that the reversible process by which stable modes return energy to the mean flow, in opposition to extraction by the instability, is quite effective at blunting the energy cascade to small scales. Stable modes have been connected to the shear-layer contraction (i.e., build up of mean flow) \cite{fraser2021}, but only as a transient process. Here we demonstrate that for a driven flow profile, the return of fluctuation energy to the mean profile occurs continuously at a rate that nearly matches the rate at which unstable modes attempt to flatten the flow profile.

The principal result of this Letter is that, in driven shear-flow MHD turbulence, the nonlinearly excited large-scale stable eigenmodes efficiently strain the magnetic fields, in a way that counteracts the effect of unstable modes, and thus greatly weakens the magnetic energy cascade to small scales.

\begin{figure}
    \centering
    \includegraphics[width=0.48\textwidth]{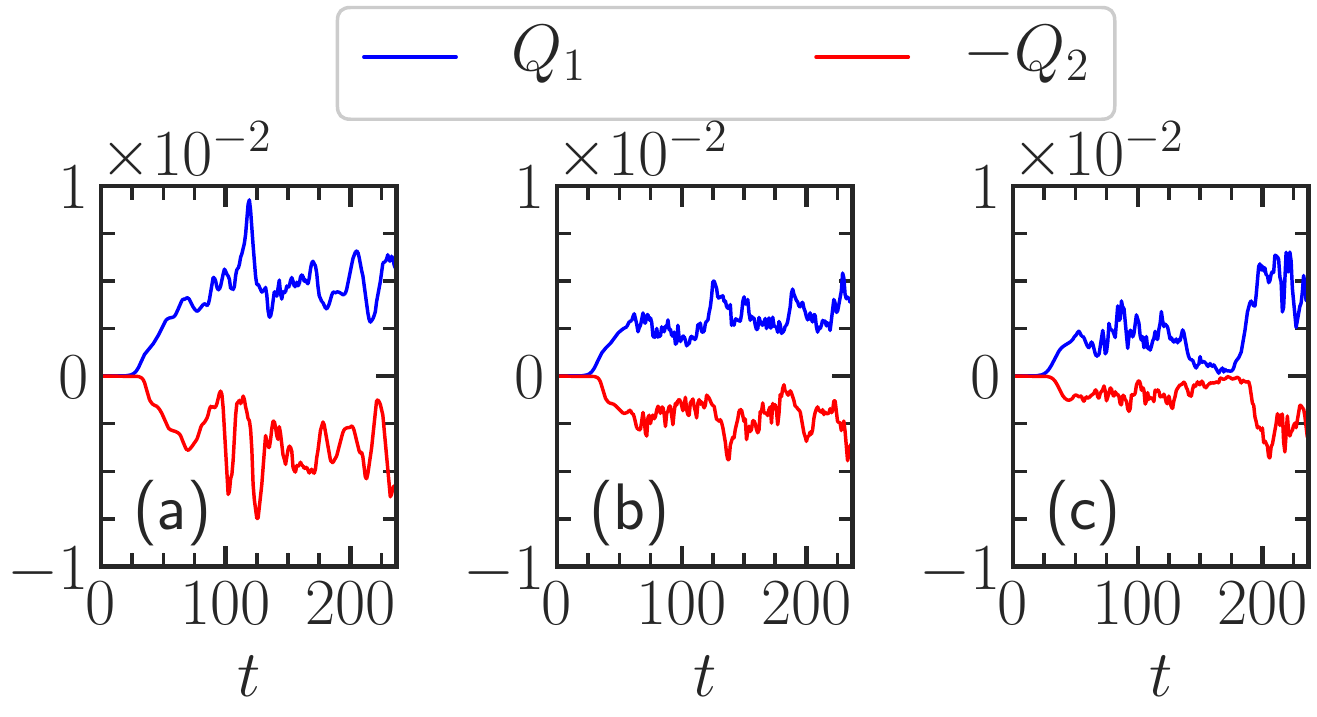}
    \caption{Time trace of $Q_1$ (blue) vs. $-Q_2$ (red) for (a) $\Pm=0.1$, $\Rm=50$; (b) $\Pm=1$, $\Rm=500$; (c) $\Pm=10$, $\Rm=5\,000$. Compare the lower (red) with upper (blue) curves within \textit{each} subplot. To illustrate similar variation of $Q_1$ and $Q_2$, and to make the variations maximally visible, we plot $-Q_2$, which physically refers to the energy transfer rate from the fluctuations to the mean profiles. Computationally demanding simulation for (c) was stopped at $t=237$. All simulations use $\mbox{M}_{\rm A}=10$. Nonlinear phase begins at $t \sim 30$. Energy available for cascading to small scales is significantly impeded by stable modes in all cases.}
    \label{fig:f2}
\end{figure}

\begin{figure*}
    \centering
    \includegraphics[width=1\textwidth]{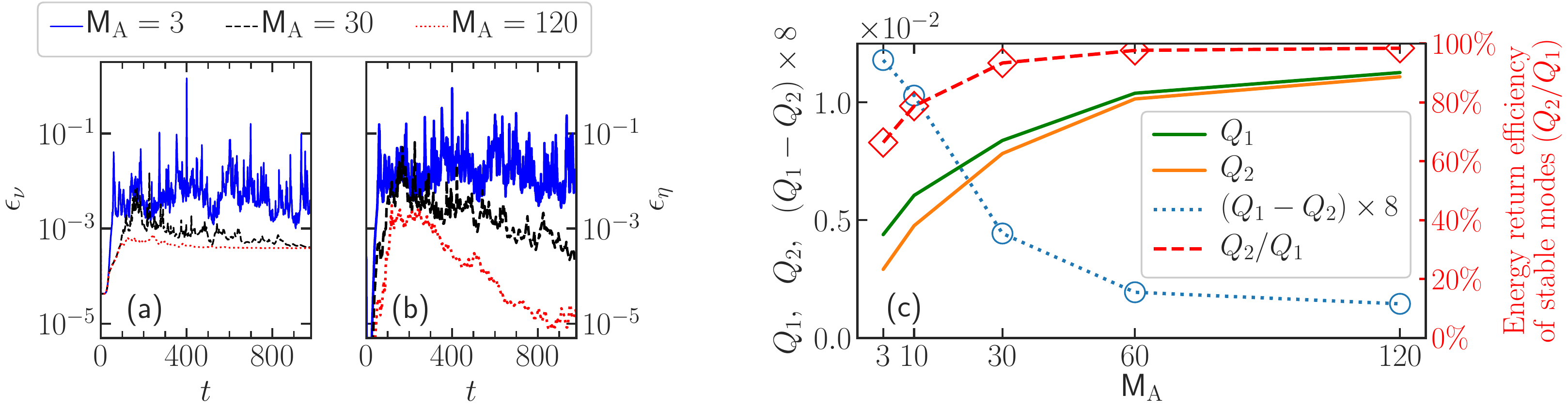}
\caption{ Time traces of (a) viscous dissipation rates, $\epsilon_\nu$,  and (b) resistive dissipation rates, $\epsilon_\eta$, for different strengths of magnetic fields. Stronger fields enhance both. (c) Impact of magnetic fields on the (time-averaged) energy transfer rates between the background profile ($k_x=0$) and the large-scale fluctuations ($k_x=0.2\textrm{--}0.8$) by the unstable eigenmodes $Q_1$ and by the stable eigenmodes $Q_2$. Note that the magnetic fields impact $Q_2$ more than $Q_1$, but $Q_2/Q_1$ is still $\gtrsim 60\%$ even for the strongest field.}
    \label{fig:f3}
\end{figure*}

We study an incompressible, two-dimensional ($2$D) system with the initial fluid velocity given as $\mathbf{v}(x,z,t=0) =U_\mathrm{ref}(z) \hat{\mathbf{x}}= U_0 \mathrm{tanh}(z/a) \hat{\mathbf{x}}$ and an initially uniform flow-aligned magnetic field as $\mathbf{B}(x,z,t=0) = B_0 \hat{\mathbf{x}}$. The parameters $a, U_0,\mathrm{ and\ } B_0$ represent the half-width of the flow-shear, the maximum fluid velocity, and the magnetic field strength, respectively, which are used to non-dimensionalize all the variables henceforth. Consequently, time and energy (per unit mass) have units of $a/U_0$ and $U_0^2$. We describe the flow and the magnetic field by a stream function $\phi$ and a flux function $\psi$, where $\mathbf{v}=\hat{\mathbf{y}} \times \nabla \phi$ and $\mathbf{B}=\hat{\mathbf{y}} \times \nabla \psi$. We study their evolution using the momentum and induction equations of the standard MHD \cite{biskamp2003, remark_supplementary}
\begin{subequations}
\begin{align} \label{eq:momentumeqn}
    &\partial_t \nabla^2 \phi =- \{\nabla^2 \phi , \phi \} + \mbox{M}_{\rm A}^{-2} \{\nabla^2 \psi , \psi \} \nonumber\\ 
    &\hspace{3.0cm}+ \Re^{-1} \nabla^4 \phi + \partial_z f(k_x\mathrm{=}0, z, t), \\ \label{eq:inductioneqn}
    &\partial_t  \psi = \{ \phi , \psi \} + \Rm^{-1} \nabla^2 \psi,
\end{align}
\end{subequations}
where $\nabla^2 \phi$ and  $\nabla^2 \psi$ are the vorticity and the current density; Poisson bracket $\{P,Q \} \equiv \partial_x P \cdot \partial_z Q - \partial_x Q \cdot \partial_z P$; the \Alfvenic\ Mach number is $\mbox{M}_{\rm A} \propto U_0/B_0$; the fluid and magnetic Reynolds numbers are defined as $\Re=U_0 a/\nu$ and $\Rm=U_0 a/\eta$.
In this study, we take $\Re=500$. Except for simulations where $\Rm=50$ and $\Rm=5~000$ (which will be stated explicitly), all others have $\Rm=500$.
In Eq.~\eqref{eq:momentumeqn}, $f$ represents a body force that continuously regenerates the mean-flow shear. The expression $k_x\mathrm{=}0$ in it indicates that the force acts only on the ($x$-averaged) mean flow, where $k_x$ is the Fourier wavenumber. The forcing thus prevents gradual flattening of the mean flow as the instability extracts energy from it. A similar process occurs in astrophysical flows, where forces like gravitation replenish continually the shear profile, e.g., shear layers in accretion disks, stars, and planetary atmospheres. We represent this force by a Krook-like operator \cite{pueschel2014, marston2008, smith2021}, sometimes referred to as a profile relaxation term \cite{allawala2020},
\begin{equation} \label{eq:forcing}
     f = D_{\mathrm{Krook}} [U_{\mathrm{ref}}(z) - \langle U(x,z,t) \rangle_x] + F_0,
\end{equation}
where $D_{\mathrm{Krook}}$ represents the rate at which the mean flow is forced towards the reference flow profile. We choose $D_\mathrm{Krook}=2$. Detailed analyses of a complete scan of different parameters will be reported elsewhere. The force $F_0$ balances the viscous diffusion of the mean shear layer, $\Re^{-1} \nabla^2 U_{\mathrm{ref}}(z)  + F_0 = 0$, in Eq.~\eqref{eq:momentumeqn}, to ensure an initial equilibrium state. 
%Now, 

We add small-amplitude perturbations to the flow and the magnetic field \cite{remark_supplementary, fraser2021} and perform time-integration of the above set of equations using the pseudospectral code Dedalus \cite{burns2020} for long times ($t> 1000$; the $\rm{e}$-folding growth time of the instability is $\approx 6$). Apart from the simulation of $\Pm=10$ that uses spectral resolution of $4096$ Fourier (Chebyshev) modes along the $x$-($z$-)axis for the box size of $(L_x, L_z) = (6\pi, 8\pi)$, all other simulations are performed at $2048 \times 2048$ spectral resolution for the box size of $(L_x, L_z) = (10\pi, 20\pi)$, while convergence checks utilize resolutions as high as $8192 \times 8192$, with no substantial impact on dissipation rates and spectral energy densities. (All simulations additionally use $3/2$ dealiasing rule \cite{burns2020}).

We also perform a corresponding eigenvalue calculation of the non-dissipative equations, derived from Eqs.~\eqref{eq:momentumeqn} \& \eqref{eq:inductioneqn}, by linearizing about the initial flow and magnetic field profiles \cite{fraser2021}. \textit{A priori}, one does not know whether such an eigenmode basis is useful to understand turbulent features. Nevertheless, since the eigenmodes of this linear operator form a complete (albeit non-orthogonal) basis \cite{remark_B, bender2019}, we can expand the time-evolving state space of nonlinear simulations in this basis after determining amplitude $\beta_j(k_x, t)$ of each eigenmode $j$ at each $k_x$ as a function of time, $\widetilde{\chi}_{\mathrm{turbulent}}(x, z, t) = \sum_{k_x:k_x \neq 0} e^{ik_x x}\left[\sum_j \beta_j(k_x, t) \chi_j(k_x, z) \right],$ where $\widetilde{\chi}_{\mathrm{turbulent}} = (\widetilde{\phi}_{\mathrm{turbulent}}, \widetilde{\psi}_{\mathrm{turbulent}})$ captures the turbulent state space and $\chi_j$ is the $j^{\mathrm{th}}$ eigenmode \cite{fraser2021, tripathi2022_sherwood, remark_energynorm, remark_ECS_PCA}. This procedure is illustrated in Fig.~\ref{fig:f1}(a) where the eigenmodes of the flow fluctuations are shown. (The magnetic fluctuations have eigenmodes~\cite{fraser2021} and parity~\cite{ishizawa2019, sato2017} related to those of the flow fluctuations.)

At each wavenumber $k_x$, linearly unstable to perturbations (i.e., $0 < k_x < 1$ for the present shear-flow instability), there is an unstable eigenmode $\phi_1$ and its conjugate stable eigenmode $\phi_2$, with eigenvalues complex conjugate to each other \cite{chandrashekhar1961, fraser2017}. All the remaining eigenmodes have purely real eigenfrequencies \cite{tripathi2022_sherwood}, belong to the continuum \cite{case1960}, and have narrow structures along the $z$-axis.

Tracking the time evolution of each eigenmode amplitude $\beta_j(k_x, t)$ in the nonlinear simulations, the initially exponentially decaying mode has been found to be nonlinearly excited to almost the same level as the unstable mode \cite{fraser2021}. The driving mechanism for this excitation is, at early times, the nonlinear interaction between the unstable modes, see, e.g., Refs.~\cite{terry2006, makwana2011, fraser2021}. In the fully nonlinear phase, however, all modes that have large energy interact dominantly.

For a given KH-unstable wavenumber, it is found, in Fig.~\ref{fig:f1}(b), that a combination of two eigenmodes from the above basis reconstructs the large-scale turbulent flow at that wavenumber. Figure~\ref{fig:f1}(b) shows a reconstruction of the turbulent flow features using unstable eigenmodes at each unstable wavenumber (left column) and with their conjugate stable eigenmodes added (right column). Simply adding one additional mode drastically decreases the difference between the true stream function and its reduced representation compared to what is achieved with a reduced representation based on the unstable modes alone (such as those predicted by quasilinear models) \cite{remark_field_reconstruction}.

Note that the stable and unstable eigenmodes of the KH linear operator, employed above, are related to one another by a space-time-reversal symmetry operation \cite{bender2019, fu2020}. Thus the unstable and stable eigenmodes transfer energy in opposite directions---the former from the mean flow to the fluctuations, while the latter in the reverse direction. The energy that is available to cascade to small scales thus depends on this competition. The magnitude of the energy transfer rates between the background shear-flow $U_\mathrm{ref}$ and the fluctuation scale $k_x \neq 0$ by the unstable and stable eigenmodes can be derived from the MHD equations \cite{remark_supplementary}: $Q_{1}(k_x) = \Gamma(k_x) |\beta_1(k_x)|^2 $ and $Q_{2}(k_x) = \Gamma(k_x) |\beta_2(k_x)|^2 $ where $\Gamma(k_x)$ is the linear growth rate of the unstable mode at that wavenumber $k_x$.

Shown in Fig.~\ref{fig:f2} are the time traces of $Q_1$ and $Q_2$ for different $\Pm$ (resistivity) over a challenging two orders of magnitude. Summation over the KH-unstable wavenumbers, $0 < k_x < 1$, is used in determining $Q_1$ and $Q_2$. Within each subplot, it is observed that $Q_1$ and $Q_2$ are nearly in equipartition in the nonlinear phase.

\begin{figure*}
    \centering
    \includegraphics[width=1\textwidth]{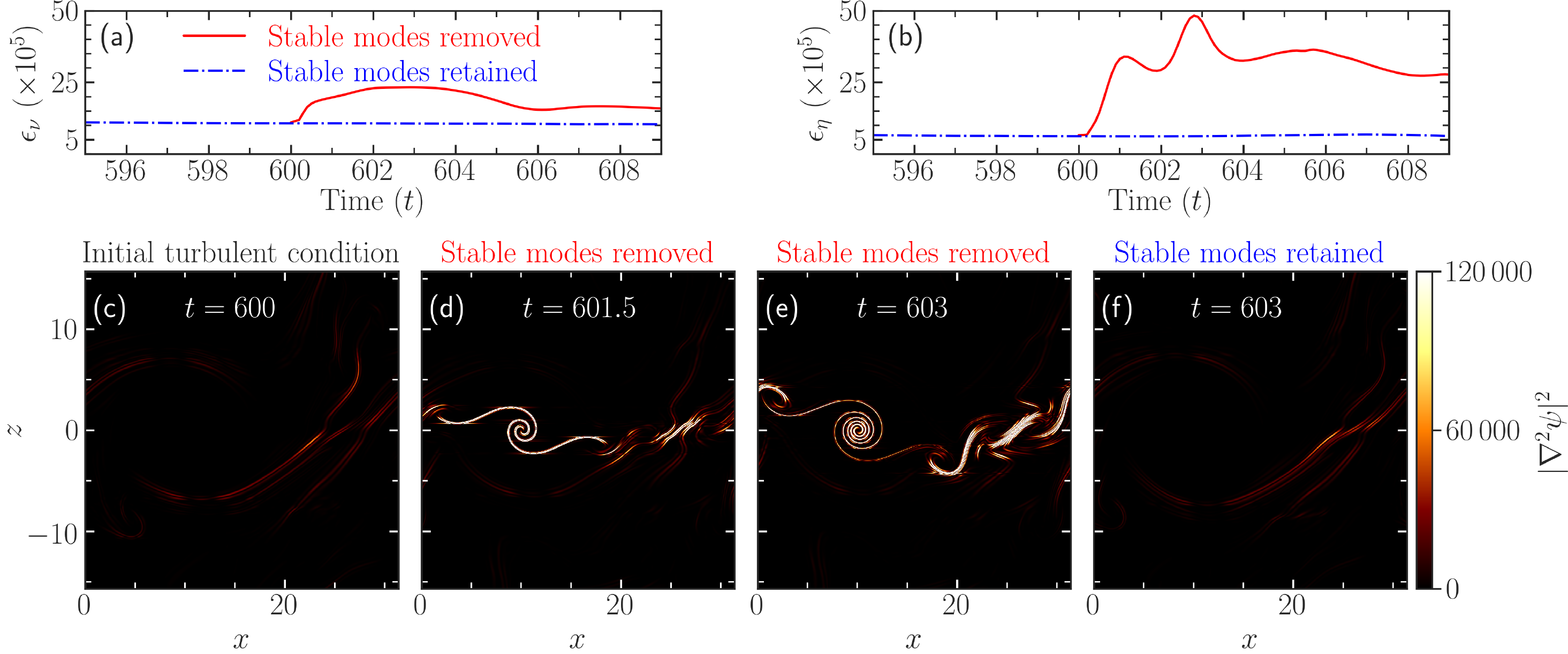}
    \caption{(a) Enhancement of the viscous dissipation rate, after removing the large-scale stable eigenmodes at an instant of time $t=600$ in a simulation with $\mbox{M}_{\rm A} = 120$. (b) Similar enhancement of the resistive dissipation. (c) The state when the simulation is paused. Shown is the squared current density at smaller scales, $|k_x|>1$, focusing on the region near the shear layer, $-5\pi \leq z \leq 5\pi$. At this instant, the large-scale stable modes belonging to $|k_x| < 1$ are deleted to observe their influence on the small-scale magnetic cascade. (d)\textrm{--}(e) Subsequent evolution of the small-scale current density when large-scale stable eigenmodes are removed at $t=600$. The characteristic enfolding of magnetic fields by the eddies of unstable modes is evident, which generates small-scale magnetic features. (f) Small-scale current density in another simulation where stable eigenmodes are kept intact. The rapid straining of magnetic fields mentioned just before does not occur. For this simulation, the state at $t=601.5$ (not shown) is almost identical to that at $t=603$. \href{https://drive.google.com/file/d/1-3lMGyba5I9A5XYe7wDwKjk6s7TeGYoP/view?usp=sharing}{(Multimedia view)}}
    \label{fig:f4}
\end{figure*}

Stable-mode-excitation process may also be impacted by magnetic fields. To analyze such, we first observe the visco-resistive dissipation with varying field strengths in Figs.~\ref{fig:f3}(a) and (b). The small-scale dissipation---a proxy of small-scale cascade---enhances with stronger fields, despite the fields reducing the linear growth rate of the instability, which sometimes is argued to lead to subdued turbulent fluctuations as the energy extraction rate by the unstable mode becomes lower.

To understand such enhanced dissipation rate, it may be instructive to compute the energy transfer rates between the background flow and the fluctuations due to both the unstable and stable eigenmodes. As seen in Fig.~\ref{fig:f3}(c), the (time-averaged) energy injection rate by the instability, $Q_1$, decreases with stronger magnetic fields, as anticipated. However, the energy return rate by the stable modes, $Q_2$, is impacted more by the stronger fields, allowing larger energy cascade to small scales ($\propto Q_1-Q_2$) that manifests as enhanced dissipation rate at such scales \cite{remark_instantaneousmean}. Thus a stronger magnetic field, until it nearly stabilizes the system, allows more net energy extraction from the free-energy source in the \textit{non}linear state. This questions models that assume the linear growth rate of an instability as a surrogate for the cascade rate or a proxy for nonlinear time, see, e.g., \cite{fuller2019, pessah2006, garaud2018, goodman1994}. Predictions based upon conventional wisdom of instability-saturation are thus challenged.

The ratio $Q_2/Q_1$ of the stable-eigenmode efficiency to the unstable-eigenmode efficiency in transferring energy between the fluctuations and the background flow is more than $60\%$ even for the strongest magnetic fields in Fig.~\ref{fig:f3}(c), see the red dashed line. Such a strong Lorentz back-reaction is most potent in disrupting the vortex \cite{mak2017} and creating more small scales, leading to more reduced $Q_2$ than the corresponding reduction in $Q_1$.

We now show visualizations of highly effective role of stable modes in structuring the magnetic fields. After carrying out an initial-value simulation up to a time when the dynamics are nonlinear and turbulent, we perform two distinct simulation continuations---one unchanged with the stable modes untouched, and another with the stable modes removed. For the latter, we set only the large-scale ($0<|k_x| < 1$) stable-eigenmode amplitudes to zero at the instant when the simulation is restarted. The differences immediately afterward are significant: Removing stable eigenmodes rapidly enhances dissipation rates, see Figs.~\ref{fig:f4}(a) and (b); \href{https://drive.google.com/file/d/1-3lMGyba5I9A5XYe7wDwKjk6s7TeGYoP/view?usp=sharing}{(Multimedia view)}. To understand this rise in the dissipation, we plot the squared current density $|\nabla^2 \psi|^2$ at small scales ($|k_x| >1$) in Figs.~\ref{fig:f4}(c)--(f). Observe in Figs.~\ref{fig:f4}(d) and (e) that the magnetic fields get rapidly distorted by the unstable modes of the flow, creating spirals as the counter-straining motion of the stable modes is removed. Note that the (un)stable modes have the largest strain around $(x,z)\approx(10,0)$ for the shown time, as can also be seen in the figure in the supplementary material. The stable modes here are deleted only at the instant when the simulation is paused; they are quickly nonlinearly driven back. In another simulation where the stable modes are untouched (retained), the spirals do not appear, however, and the time evolution immediately afterward remains almost identical to the initial stage [cf. Figs.~\ref{fig:f4}(f) and Fig.~\ref{fig:f4}(c)]. This exercise of instantaneous deletion of stable modes exposes the true nonlinear impact of the stable modes on magnetic field evolution, which remained hidden in Fig.~\ref{fig:f1}(a).

This numerical experiment illustrates that the energy injected by the unstable modes into the large-scale fluctuations ($|k_x| <1$) via mean-fluctuation interactions, \textit{in the absence of stable modes}, cascades in its entirety to smaller scales nonlinearly where it dissipates visco-resistively. This is a statement of the classical small-scale cascade by the advective nonlinearity, $(\mathbf{v}\cdot \bm{\nabla})\psi$, as commonly understood. When $\mathbf{v}$ is composed of comparable unstable and stable mode-amplitudes, though, the cascade is weakened by a factor of up to ten [Figs.~\ref{fig:f2},~\ref{fig:f3}(c),~and~\ref{fig:f4}(b)]. Such a finding informs and can be used to improve reduced models of geo- and astrophysical instability-driven turbulence \cite{vandine2021, fuller2019}, where all the energy injected by the instability is assumed to pass through an inertial cascade, e.g., Refs.~\cite{fuller2019, pessah2006, garaud2018, goodman1994}.

It may be noted that, although $2$D and $3$D turbulence have different conserved quantities and corresponding cascade processes, the stable modes of this study are the conjugate roots of the KH-instability, which exist in both $2$D and $3$D \cite{chandrashekhar1961}. Our preliminary studies of $3$D turbulence have indeed shown that virtually identical stable mode excitation and its impact in turbulence occur in $3$D as well. Details of such an investigation will be left for a future publication \cite{tripathi2022_testingQL}.

This Letter demonstrates, for the first time, that stable modes via their counteracting straining motion of the flow, sequester magnetic energy at large scales, despite the conventional association of shear-flow with intensification and generation of small scales in MHD turbulence. Our analyses show that the stable modes greatly weaken the magnetic energy cascade to small scales and hence can thwart the magnetic-energy accumulation at sub-viscous scales in the high-magnetic-Prandtl-number regime, typically encountered in diffuse astrophysical plasmas; thus possibly allowing operation of a dynamo with magnetic fields concentrated at large scales, in accord with astrophysical observations \cite{schekochihin2002,tobias2013}. This work opens a new direction towards such studies. 

\section*{Supplementary Material}
See the supplementary material for the details of (I) simulation set-up, (II) modal energy evolution equation, (III) energy transfer rates between the background flow and the fluctuations, and (IV) residuals in low-order representation of the turbulent flow in Fig.~\ref{fig:f1}(b). A \href{https://drive.google.com/file/d/1-3lMGyba5I9A5XYe7wDwKjk6s7TeGYoP/view?usp=sharing}{simulation movie} supplements Fig.~\ref{fig:f4} of this Letter.

\section*{Acknowledgements}
\begin{acknowledgments}
The authors thank K.~Burns and Dedalus developers for assistance. Thanks also to N.~Hurst for offering helpful discussions and a careful reading of the manuscript. This material is based upon work funded by the U.S. Department of Energy [DE-SC0022257] through the NSF/DOE Partnership in Basic Plasma Science and Engineering.
We also gratefully acknowledge the Award no.~DE-FG02-04ER54742, supported by the U.S. DOE, and Grant Nos.~AST-1814327 and AST-1908338 supported by the U.S. NSF. All simulations reported herein were performed using the XSEDE supercomputing resources via Allocation No.~TG-PHY130027. 

The data that support the findings of this study are available from the corresponding author upon reasonable request.
\end{acknowledgments}


\begin{thebibliography}{9}
\bibitem{remark_energynorm}
Energy norm is used to normalize the eigenmodes.
\bibitem{remark_field_reconstruction}
Thus reconstructed flow includes components that strain the magnetic field efficiently. The turbulent magnetic field, in contrast, requires many eigenmodes for each wavenumber. This is consistent with the fluid-straining-induced generation of small scales of the magnetic field, which when projected onto the eigenmode space covers a large number of marginally-stable continuum eigenmodes.
\end{thebibliography}
\end{document}

%\preprint{APS/123-QED}

%\showthe\textwidth %I have found in the log file control+find button for \showthe\textwidth yields 510 pts. [BINDESH TRIPATHI, AUG 08, 2021]

%\title{Nonlinear saturation of MHD shear-flow instability and impeded rapid distortion of magnetic fields by stable eigenmodes }
\title{Supplementary materials for ``Mechanism for Sequestering Magnetic Energy at Large Scales in Shear-Flow Turbulence"}

\author{B.~Tripathi}
\affiliation{University of Wisconsin-Madison, Madison, Wisconsin 53706, U.S.A.}
\author{A.E.~Fraser}
\affiliation{University of California, Santa Cruz, Santa Cruz, California 95064, U.S.A.}
\author{P.W.~Terry}
\author{E.G.~Zweibel}
\affiliation{University of Wisconsin-Madison, Madison, Wisconsin 53706, U.S.A.}

\author{M.J.~Pueschel}
\affiliation{Dutch Institute for Fundamental Energy Research, 5612 AJ Eindhoven, The Netherlands}
\affiliation{Eindhoven University of Technology, 5600 MB Eindhoven, The Netherlands}

%\date{\today}

\maketitle

This document provides supplemental information related to the article ``Mechanism for Sequestering Magnetic Energy at Large Scales in Shear-Flow Turbulence." 
Section~I details the simulation set-up. Section~II presents the evolution equation for the modal energy, whenceforth an expression for the energy transfer between the background flow and the fluctuations is derived in Sec.~III, which is numerically computed and plotted in Figs.~2 and 3 of the main article. In Sec.~IV, residuals in low-order representation of the turbulent flow are compared.

\section{I. Simulation set-up and details}
We consider standard incompressible magnetohydrodynamic (MHD) equations

\begin{subequations}
\begin{align}
    &\mathbf{\nabla} \cdot \mathbf{u}=0,\\  \label{eq:mhdu}
    &\partial_t \mathbf{u} + \mathbf{u} \cdot \mathbf{\nabla} \mathbf{u} = -\frac{\nabla P}{\rho} + \frac{\left(\nabla \times \mathbf{B}\right) \times \mathbf{B}}{4\pi \rho} + \nu \nabla^2 \mathbf{u} + \mathbf{f},\\
    &\mathbf{\nabla} \cdot \mathbf{B}=0,\\  \label{eq:mhdb}
    &\partial_t  \mathbf{B} = \nabla \times \left(\mathbf{u} \times \mathbf{B} \right) + \eta \nabla^2 \mathbf{B},
\end{align}
\end{subequations}
where $\mathbf{u}$, $\mathbf{B}$, $P$, $\rho$, $\nu$, $\eta$, $\mathbf{f}$ represent the fluid velocity, magnetic field, pressure, fluid density, viscosity, resistivity, and force per unit mass of the magneto-fluid respectively. Henceforth, we confine ourselves to a two-dimensional $(x,z)$ system with the initial fluid velocity given as $\mathbf{v}(x,z,t=0) =U_\mathrm{ref}(z) \hat{\mathbf{x}}= U_0 \mathrm{tanh}(z/a) \hat{\mathbf{x}}$ and an initially-uniform flow-aligned magnetic field as $
\mathbf{B}(x,z,t=0) = B_0 \hat{\mathbf{x}} $. The parameters $a, U_0,\mathrm{ and\ } B_0$ represent the half-width of the flow-shear, maximum fluid velocity, and magnetic field, respectively. These parameters are utilized to non-dimensionalize all the variables and calculations henceforward. 

The two-dimensional system allows us to employ a more convenient and economical formalism, using stream function ($\phi$) and flux function ($\psi$). Define $\mathbf{u}= \mathbf{\hat{{y}}} \times \nabla \phi$ and $\mathbf{B}= \mathbf{\hat{{y}}} \times \nabla \psi$ so that the vorticity is $\nabla^2 \phi$ and the current density is $ \nabla^2 \psi$. Taking the curl of Eq.~\eqref{eq:mhdu} and rewriting Eq.~\eqref{eq:mhdb} in terms of the flux function, we arrive at (with all quantities non-dimensionalized with respect to $U_0$, $a$, $\mbox{M}_{\rm A} \propto U_0/B_0$)
\begin{subequations}
\begin{align} \label{eq:momentumeqn}
    \begin{split}
    &\partial_t \nabla^2 \phi + \{\nabla^2 \phi , \phi \} = \mbox{M}_{\rm A}^{-2} \{\nabla^2 \psi , \psi \} + \Re^{-1} \nabla^4 \phi +  \partial_z f(k_x\mathrm{=}0, z, t),
    \end{split}\\ \label{eq:inductioneqn}
    &\partial_t  \psi = \{ \phi , \psi \} + \Rm^{-1} \nabla^2 \psi,
\end{align}
\end{subequations}
where the external body force is considered to be $\mathbf{f} = f(k_x\mathrm{=}0, z, t)\ \mathbf{\hat{x}}$, which acts on the instantaneous mean flow and is directed along the $x$-axis. Here, $k_x$ stands for the Fourier wavenumber.

These equations are solved using the pseudospectral code Dedalus, with dealiasing according to the standard $3/2$ rule. Explicit viscosity and resistivity are used throughout all simulations. The fluid and magnetic Reynolds numbers are defined as $\Re=U_0 a/\nu$ and $\Rm=U_0 a/\eta$, with shear-width $a$ as the characteristic length scale. Throughout this study, we consider $\Re=\Rm=500$, except for two cases where $\Rm$ is changed to $\Rm=50$ and $\Rm=5\,000$, which are explicitly mentioned in Fig.~2 of the main article.

The simulation box is of size $(L_x, L_z) = (10 \pi, 20\pi)$, unlike $(10 \pi, 10\pi)$ in a previous study \cite{fraser2021}. The box is considered taller along the $z$-axis to keep any potential effect of the boundary on turbulent dynamics as minimal as possible. The vortices in the simulations are observed to be located near the shear layer and thus far from the boundaries. The simulations are then performed using the least spectral modes/resolution of $(N_x, N_z) = (2048, 2048)$. Note that Dedalus uses spectral basis to solve the nonlinear partial differential equations \cite{burns2020}. We employ Fourier modes along the $x$-axis and Chebyshev polynomials along the $z$-axis. The smallest Fourier wavenumber in our simulations is $k_x=2\pi/L_x = 0.2$. The boundary conditions used are periodic in $x$ and for the $z$-axis, perfectly conducting, no slip, co-moving [with the initial fluid flow, $\mathbf{v}(z,t=0)$] walls at $z=\pm L_z/2$ are used as in Ref. \cite{fraser2021}.

We use the following forms of the initial perturbations to start the simulations, by exciting all Fourier wavenumbers as a generalization \cite{fraser2021} to many routinely studied form of initial perturbations that excite only a particular or a few Fourier wavenumbers
\begin{equation}
\begin{aligned}
    \phi(x,z,t=0)
    &= A_\phi \sum_{k_x \neq 0} k_x^a e^{i r_\phi(k_x)} e^{-z^2/\sigma^2} e^{ik_x x},
\end{aligned}
\end{equation}
and
\begin{equation}
    \psi(x,z,t=0) = A_\psi \sum_{k_x \neq 0} k_x^a e^{i r_\psi(k_x)} e^{-z^2/\sigma^2} e^{ik_x x}.
\end{equation}
Here, $A_\phi$ and $A_\psi$ set the overall amplitude of the perturbations for the stream function ($\phi$) and flux function ($\psi$); $a$ dictates the steepness of the energy spectrum in the initial perturbations; $\sigma$ determines the width of the Gaussian profile of the perturbation in the $z$-axis; $r_\phi(k_x)$ and $r_\psi(k_x)$, defined at each Fourier wavenumber $k_x$, are different random phases, selected from a pesudo-random number generator, which are uniformly distributed between $[0, 2\pi)$. For the simulations reported here, $a=-1$, $\sigma=2$, and $A_\phi=A_\psi = 10^{-3} $ were considered, allowing a markedly defined linear evolution phase of the instability.

\section{II. Modal energy evolution equations}

Here, we derive rates at which \textit{each eigenmode} gains or gives away energy at any instant in time. Denoting this energy transfer rate by $\dot{E}_j(k_x)$ where $j$ represents an arbitrary eigenmode at the Fourier wavenumber $k_x$, we may write an expression for the energy evolution rate (the energy spectrum is symmetric across $k_x=0$ and hence the $-k_x$ and $+k_x$ have the same energy)
\begin{align}
    \dot{E}_j(k_x) {}&= \frac{dE_j^{ \mathrm{Kinetic}+\mathrm{Magnetic}}(k_x)}{dt} \nonumber\\
    &=\frac{1}{2}\frac{d}{dt} \left[ \frac{1}{L_z}\int dz \left( \frac{|\mathbf{u}_{j}(k_x)|^2}{2} + \frac{|\mathbf{B}_j(k_x)|^2}{2 \mbox{M}_{\rm A}^2}\right)\right]
    %extra 1/2 in the beginning here because when we take 1/2u^{-k_x}*u^{k_x}, this represents the total sum of energy in both negatve and positive kx spectrum. If I want to make Q_j = Gamma_j |beta_j|^2 n_j with n_j being exactly the energy in the jth eigenmode, I should introduce a factor of 1/2 and study only the positive k_x spectrum... So,my plot for fig 3 represents this. only one half of the spectrum.Since the spoectrum is symmetric across k_x=0, it is justifiable to use one of the conventient forms and make the equations on the PRL main doc familiar and easy to read.
    \nonumber\\ \label{eq:nearlylast}
    &= \frac{1}{4L_z}\frac{d}{dt}\left[\int dz \left\{|\beta_j i k_x \phi_j|^2 + | \beta_j \partial_z \phi_j|^2 +  \frac{1}{\mbox{M}_{\rm A}^2}\left(|\beta_j i k_x \psi_j|^2 + |\beta_j \partial_z \psi_j|^2\right) \right\}\right],
\end{align}
where $\beta_j =\beta_j(k_x)$ is the amplitude of the $j^{\mathrm{th}}$ eigenmode, i.e., $\left[\phi_j, \psi_j\right]=\left[\phi_j(k_x), \psi_j(k_x)\right]$ at the Fourier wavenumber $k_x$. The last equality of Eq.~\eqref{eq:nearlylast} can be rewritten in a more insightful way as
\begin{equation}
    \label{eq:lastmode}
    \dot{E}_j(k_x)= \frac{1}{2}\frac{d|\beta_j(k_x)|^2}{dt} n_{j},
\end{equation}
with
\begin{equation}
    n_{j} = \frac{1}{2L_z} \int dz \left\{|k_x \phi_j|^2 + |\partial_z \phi_j|^2 + \frac{1}{\mbox{M}_{\rm A}^2}\left(|k_x \psi_j|^2 + |\partial_z \psi_j|^2 \right)  \right\}.
\end{equation}
Here, $n_j$ represents the amount of total energy residing in each normalized eigenmode (or eigenfunction). If the energy norm is taken to normalize the eigenmodes, this quantity is automatically \textbf{unity}. So,
\begin{equation}
n_j = 1.
\end{equation}
If another norm like $L_2$ is employed, this quantity has to be evaluated separately. Note, however, that the energy transfer rate, $Q_j(k_x)$, is independent of the choice of the norm. This is because one can switch from an arbitrary norm to the energy norm by dividing the eigenfunctions by the ratio, $r=\sqrt{n_{j} \mathrm{\ in \ arbitrary\ norm}/{n_{j} \mathrm{\ in\ energy\ norm}}} = \sqrt{{n_{j} \mathrm{\ in \ arbitrary\ norm}}}$ and multiplying the eigenmode amplitudes in the eigenmode expansion by the same ratio. For example, the eigenmode expansion, defined in a basis, $\chi_j = (\phi_j, \psi_j)$, with an arbitrary norm can be written as $\chi_{\mathrm{any \ fluctuation}} = \beta_1 \chi_1 + \beta_2 \chi_2 + ... = 2\beta_1 (\chi_1/2) + 2\beta_2 (\chi_2/2) + ... = 3\beta_1 (\chi_1/3) + 3\beta_2 (\chi_2/3) + ... = r\beta_1 (\chi_1/r) + r\beta_2 (\chi_2/r) + ... $ where $r$ was defined just before and the last expansion, $\chi_{\mathrm{any \ fluctuation}} = r\beta_1 (\chi_1/r) + r\beta_2 (\chi_2/r) + ... $ , is the one that we encounter in the eigenmode expansion with the energy norm. Alternatively, $\chi_j \mathrm{\ in\ the\ energy\ norm} = \chi_j \mathrm{\ in\ an\ arbitrary\ norm}/r$, and $\beta_j \mathrm{\ in\ the\ energy\ norm} = r \times \beta_j \mathrm{\ in\ an\ arbitrary\ norm}$. In Eq. \eqref{eq:lastmode} above, the normalization factor $r$, however, cancels out as the physical transfer of energy should be independent of the normalization used. Equation \eqref{eq:lastmode} requires a knowledge of how the eigenmode amplitude $|\beta_j|$ evolves with time. This portion will be achieved in the next section.

\section{III. Energy transfer rates between the background flow and the fluctuations}
The governing nonlinear partial differential equations, which are the MHD equations for our case, can be cast onto their linear eigenmode basis from the physical $(x,z)$-space. This is a mere change of basis and one, therefore, obtains an equivalent set of dynamical equations. Details on how to do so in simplified problems can be found in Refs.~\cite{terry2006, fraser2017, terry2018}. This same procedure has been much utilized in many problems of homogeneous turbulence including homogeneous MHD turbulence where Elsasser variables are realized as linear eigenmodes of the system. Techniques similar to this were also applied in nonlinear tidal instabilities \cite{burns2018}. For quadratic nonlinearity with inhomogeneity along the $z$-axis, the evolution equation for the eigenmode amplitude, $\beta_j(k_x)$, assumes the following general form

\begin{equation}  \label{eq:threewave}
   \partial_t \beta_j(k_x) = i \omega_j(k_x) \beta_j(k_x) + \sum_l D_{jl}(k_x)\beta_l(k_x) + \sum_{\substack{k_x', k_x'', l, m\\ k_x-k_x'-k_x''=0}} C_{jlm}(k_x, k_x') \beta_{l}(k_x') \beta_{m}(k_x''),
\end{equation}
where $k_x-k_x'-k_x''=0$ is the selection rule for the three Fourier wavenumber interaction; $\omega_j(k_x)$ is the linear eigenfrequency for eigenmode $j$ at Fourier wavenumber $k_x$, or in other words, the coupling of the fluctuation with the initial mean profile. Since the instantaneous mean profiles are not exactly the same as the initial mean profiles, despite the flow being forced, a tiny linear coupling $D_{jl}(k_x)$ arises between the eigenmodes $j$ and $l$ at wavenumber $k_x$ via the interaction with the fluctuating component of the mean profiles. The effect of the viscous and the resistive dissipation are also captured by this small linear coupling, as the eigenmodes used here are the eigenmodes of the non-dissipative linear operator. Despite this linear coupling, we have computed and found that its contribution in the energy transfer is small compared to the coupling via the initial mean profiles, represented by $\omega_j(k_x)$. Next, the nonlinear coupling strength of fluctuations is represented by $C_{jlm}(k_x, k_x')$. This coupling allows two eigenmodes ($l, m$) at two different Fourier wavenumbers ($k_x', k_x''$) to beat together to drive an eigenmode ($j$) at another Fourier wavenumber ($k_x$). We have computed the contribution from this nonlinear mode-coupling also, and found, as expected, that the time-averaged nonlinear energy transfer rate to any wavenumber $k_x$ is equal, but opposite in sign, to the time-averaged linear injection rate to the same wavenumber $k_x$.

Multiplying Eq.~\eqref{eq:threewave} with $\beta_j^\ast(k_x)$, the complex conjugate of $\beta_j(k_x)$, and summing the resulting equation with its complex conjugate, an equation for the time evolution of $|\beta_j(k_x)|^2$ is obtained. This equation is then substituted in Eq.~\eqref{eq:lastmode} as shown below, with $n_j=1$ \cite{makwana2012}
\begin{align}
    \dot{E}_j(k_x)&= \frac{1}{2}\left[ \beta_j^{\ast} \partial_t \beta_j + \beta_j \partial_t \beta_j^{\ast} \right]\nonumber\\
    &=  \mathrm{Re} \left[ \beta_j^{\ast} \partial_t \beta_j\right] \nonumber\\
    &= \mathrm{Re} \left[ \beta_j^{\ast}  i \omega_j \beta_j + \beta_j^{\ast}  \sum_l D_{jl}(k_x)\beta_l + \beta_j^{\ast} \sum_{\substack{k_x', k_x'', l, m\\ k_x-k_x'-k_x''=0}} C_{jlm}(k_x, k_x') \beta_{l}' \beta_{m}'' \right] \nonumber\\
    &= \mathrm{Re} \left[  \Gamma_j |\beta_j|^2 + \sum_l D_{jl}(k_x)\beta_j^{\ast} \beta_l  +  \sum_{\substack{k_x', k_x'', l, m\\ k_x-k_x'-k_x''=0}} C_{jlm}(k_x, k_x') \beta_{l}' \beta_{m}''\beta_j^{\ast} \right] \nonumber\\ \label{eq:modeenergy}
    &= \Gamma_j |\beta_j|^2 +  \mathrm{Re}\left[ \sum_l D_{jl}(k_x)\beta_j^{\ast} \beta_l +  \sum_{\substack{k_x', k_x'', l, m\\ k_x-k_x'-k_x''=0}} C_{jlm}(k_x, k_x') \beta_{l}' \beta_{m}''\beta_j^{\ast}  \right],
\end{align}
where $\beta_{l}'$ and $\beta_{m}''$ represent $\beta_{l}(k_x')$ and $\beta_{m}(k_x'')$ respectively. Unprimed quantities are to be evaluated at $k_x$. The quantity $i \omega_j$ has been replaced by $\Gamma_j$ as it is the growth rate for the unstable or stable eigenmode (i.e., $e^{i \omega_j t} = e^{\Gamma_j t}$). It may be noted that each of the terms in Eq.~\eqref{eq:modeenergy} has been evaluated. A detailed analysis of those energy transfer channels and mechanisms will be reported in a separate forthcoming publication.

Since the free energy source is the background shear flow in this study, the first term on the right hand side of the last equality in Eq.~\eqref{eq:modeenergy} is of interest to compute the direct energy transfer rate between the background and the fluctuations. This term informs us about how much energy can cascade down to small scales from the free energy source at any instant in time. 

The equation
\begin{equation}
    Q_j(k_x) = \Gamma_j |\beta_j|^2 ,
\end{equation}
is precisely the one that is presented in the main article. Note that $Q_j(k_x)$ represents the energy transfer rate between the background profiles and the fluctuations due to the $j^{\mathrm{th}}$ eigenmode at the wavenumber $k_x$, unlike $\dot{E}_j(k_x)$ that adds up all the effect of the transfer from the background to the fluctuations, the re-distribution among the fluctuation scales via three-wave interactions, and the visco-resistive dissipation \cite{makwana2012}. Since the quantity $Q_j(k_x)$ measures the rate at which the energy gets depleted from the background flow and cascades to small scales, this quantity is numerically evaluated during post-processing of the direct numerical simulation data, and their time-averages (for the turbulent phase $t=350\textrm{--}800$) are plotted in Fig.~3(c) of the main article.
\begin{figure*}
    \centering
    \includegraphics[width=1\textwidth]{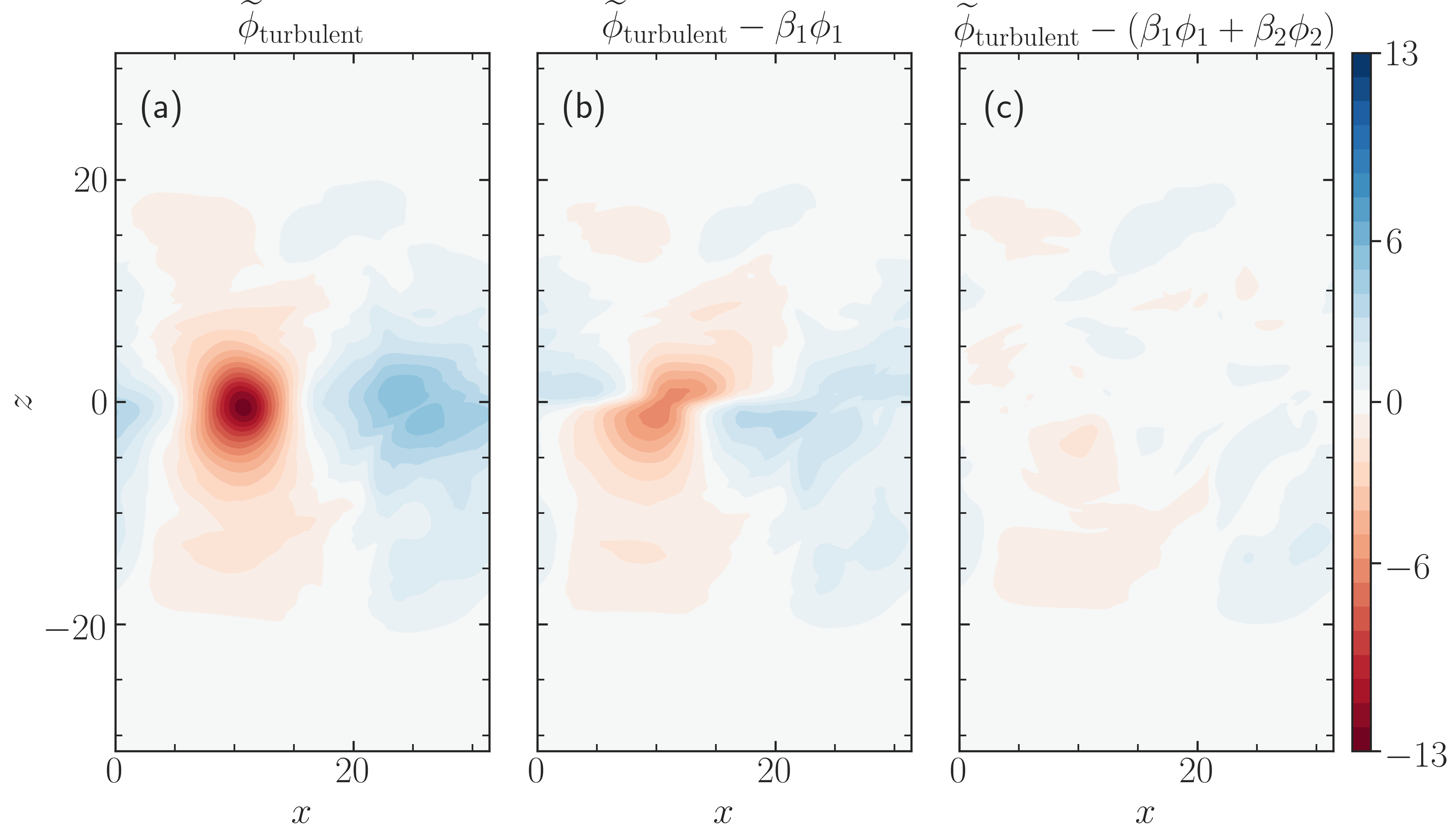}
    \caption{(a) Full turbulent fluctuations $\widetilde{\phi}_\mathrm{turbulent}$, as shown in Fig.~1(a) of the main article. The aspect ratio of the figure is restored here so that the actual box of the simulation and the turbulent features therein can be seen in their true sizes. (b) The residual $\widetilde{\phi}_\mathrm{turbulent} - \beta_1 \phi_1$ after subtracting the fluctuations due to unstable eigenmodes. It can be seen that the this plot appears very similar to the stable mode, shown in Fig.~1(a) of the main article. (c) Dramatic reduction in the residual $\widetilde{\phi}_\mathrm{turbulent} - (\beta_1 \phi_1 + \beta_2 \phi_2)$ by adding a conjugate stable mode to its unstable mode at each Kelvin-Helmholtz-unstable wavenumber. The small residual observed in this plot belongs to a large number of continuum eigenmodes, which have their mode structures somewhat afar from the shear layer $|z| \lesssim 1$. All panels share the same colorbar.}
    \label{fig:f1}
\end{figure*}

% The source codes for the calculations performed in this study, along with the numerical data, are available from the authors upon reasonable request. %Email correspondence are warmly welcome to btripathi@wisc.edu, adfraser@ucsc.edu, pwterry@wisc.edu or any of the other co-authors.

\section{IV. Residuals in low-order representation of the turbulent flow}

It can be insightful to analyze the residuals in Fig.~1(b) of the main article, in order to learn about the turbulent features that are captured by the unstable modes alone and by the sum of the unstable and stable modes. Such an analysis is shown here in Fig.~1.

Figure~1 here also helps understand why spirals form in Fig.~4 of the main article. The stable mode structure observed in Fig.~1(b) here, when removed, loses the ability to counteract the unstable mode in straining the magnetic fields. Thus the ensuing straining and folding of the field by the unbridled eddy motion of the unstable mode gives rise to spirals, beginning at around $(x,z)\approx (10, 0) $---the center or ``eye" of the stable-mode-eddy. (Note that the location of this center evolves over a longer time.)